\documentclass[a4paper,twoside,twocolumn,english,aps,sort,amsfonts,prb,hyperref,floatfix,preprintnumbers,showpacs,showkeys]{revtex4}
\usepackage[T1]{fontenc}
\usepackage[latin1]{inputenc}
\usepackage{amsmath}
\usepackage{babel}
\usepackage{graphics}
\usepackage{amssymb}
\usepackage{color}
\makeatletter

\providecommand{\LyX}{L\kern-.1667em\lower.25em\hbox{Y}\kern-.125emX\@}
\let\SF@@footnote\footnote
\def\footnote{\ifx\protect\@typeset@protect
    \expandafter\SF@@footnote
  \else
    \expandafter\SF@gobble@opt
  \fi
}
\expandafter\def\csname SF@gobble@opt \endcsname{\@ifnextchar[
  \SF@gobble@twobracket
  \@gobble
}
\edef\SF@gobble@opt{\noexpand\protect
  \expandafter\noexpand\csname SF@gobble@opt \endcsname}
\def\SF@gobble@twobracket[#1]#2{}

\makeatother
\begin{document}

\preprint{preprint to be submitted to Phys. Rev.}

\title{MgCNi\( _{3} \) : Complex behavior in a strongly coupled
superconductor}

\author{A. Wälte$^{\dag}$, 
H. Rosner$^{\ddag}$,
M.D. Johannes$^{\ddag}$,
G. Fuchs$^{\dag}$
\email{fuchs@ifw-dresden.de (G.Fuchs)}
K.-H. Müller$^{\dag}$,
A. Handstein$^{\dag}$,
K. Nenkov$^{\dag}$,
\thanks{On leave from: Int Lab. of High Magn. Fields, Wroclaw; ISSP-BAS,
Sofia, Bulgaria.}
V.N. Narozhnyi$^{\dag}$,
\thanks{On leave from Inst. for High Pressure Physics, Troitsk, Russia.}
S.-L. Drechsler$^{\dag}$,
S. Shulga$^{\dag}$, and
L. Schultz$^{\dag}$
}

\affiliation{$^{\dag}$ Institut für Festkörper- und Werkstoffforschung
Dresden, Postfach 270116, D-01171 Dresden, Germany\\}
\affiliation{$^{\ddag}$Department of Physics, University of California, Davis, California
95616}

\date{1 August 2002}

\begin{abstract}
Polycrystalline samples of the recently discovered MgCNi\( _{3} \)
superconductor were investigated by transport, ac susceptibility,
dc magnetization and specific heat measurements in magnetic fields
up to \( 16\textrm{T} \). Consistent results were obtained for the
temperature dependence of the upper critical field \( H_{\textrm{c}2}(T) \)
from resistance, ac susceptibility and specific heat measurements.
A WHH like temperature dependence of  \( H_{\textrm{c}2}(T) \)
and the quadratic relationship \( H_{\textrm{c}2}(0)\sim T_{\textrm{c}}^{2} \)
point to an effective predominant single band behavior near the quasi
clean limit. Evidence for strong electron-phonon and electron-paramagnon
coupling was found analyzing the specific heat data. The corresponding
s- and p-wave scenarios are briefly discussed using calculated densities
of states of different Fermi surface sheets.
\end{abstract}



\maketitle

\section{Introduction}

The recent discovery of superconductivity in the intermetallic antiperovskite
compound MgCNi\( _{3} \)\cite{he01} with a superconducting transition
temperature of \( T_{\textrm{c}}\simeq 8\textrm{ K} \) is rather
surprising considering its high Ni content. Therefore, it is
not unexpected that this compound is near a ferromagnetic instability
which might be reached by hole doping at the Mg sites.\cite{rosner02}
The possibility of unconventional superconductivity due to the proximity
of these two types of collective order has attracted great interest
in the electronic structure and the physics of the pairing mechanism. Band
structure calculations\cite{rosner02,singh, jarlborg} for MgCNi\( _{3} \) reveal a domination
of the electronic states at the Fermi surface by the 3d orbitals of
Ni.

MgCNi\( _{3} \) can be considered as the 3-dimensional analogue of
the layered transition metal borocarbides which have superconducting transition
temperatures up to \( T_{\textrm{c}}\simeq 23\textrm{ K} \). In spite
of the much lower \( T_{\textrm{c}} \) of MgCNi\( _{3} \), its upper
critical fields \( H_{\textrm{c}2} \) at low temperatures is, with
\( H_{\textrm{c}2}(0)=10\ldots 15\textrm{ T} \)\cite{li,li01,mao,lin}
comparable with that of the borocarbides or even higher. This is
connected with the completely different temperature dependence of
\( H_{\textrm{c}2} \) for these compounds. The \( H_{\textrm{c}2}(T) \)
dependence of MgCNi\( _{3} \) is similar to that of dirty-limit
intermetallic superconductors with a steep slope of \( H_{\textrm{c}2}(T) \)
at \( T_{\textrm{c}} \).

Through analysis of specific heat data, MgCNi\( _{3} \) was
characterized in the framework of a conventional, phonon-mediated
pairing both as a moderate \cite{he01,lin} and as a strong\cite{mao}
coupling superconductor.  Strong coupling is also suggested by the large
energy gap determined from tunneling experiments.\cite{mao}
The question of the pairing symmetry is controversially discussed in
the literature.\(
^{13} \)C NMR experiments\cite{singer01} support s--wave pairing in
MgCNi\( _{3} \), whereas tunneling spectra indicate an unconventional
pairing state.\cite{mao}

In the present investigation, upper critical field and specific heat
data of MgCNi\( _{3} \) were analyzed with the aid of theoretical
results for Fermi velocities and partial densities of states in order
to find out a consistent physical picture explaining the experimental
results.

\section{Experimental}

\begin{figure}[tbph]
{\centering \resizebox*{8cm}{6cm}{\rotatebox{-90}{\includegraphics{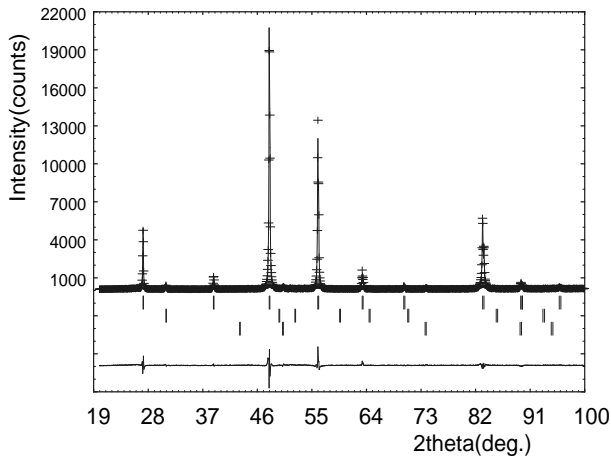}}} \par}

\caption{Rietveld refinement for MgC\(_{1.6}\)Ni\protect\( _{3}\protect \). The crosses
correspond to the experimental data. The solid line shows the calculated
pattern. Vertical bars give the Bragg positions for the main phase
MgCNi\protect\( _{3}\protect \), graphite and MgO (from top to
bottom). The line at the bottom gives the difference
between the experimental and calculated data. \label{xray Diffraktogramm}}
\end{figure}

Polycrystalline samples of MgCNi\( _{3} \) have been prepared by
solid state reaction. In order to obtain samples with high \( T_{\textrm{c}} \),
one has to use an excess of carbon as stated in Ref. \onlinecite{he01}.
To cover the high volatility of Mg during the sintering of the samples
an excess of Mg is needed, too.\cite{he01} In this study, a sample
with the nominal formula Mg\( _{1.2} \)C\( _{1.6} \)Ni\( _{3} \)
has been investigated which is denoted as MgC\( _{1.6} \)Ni\( _{3} \).
To prepare the sample, a mixture of Mg, C and 
Ni powders was pressed into a pellet. The pellet was wrapped in a
Ta foil and sealed in a quartz ampoule containing an Ar atmosphere
at 180 mbar. The sample was sintered for half an hour at \( 600^{\circ }\textrm{C} \)
followed by one hour at \( 900^{\circ }\textrm{C} \). After a cooling
process the sample was regrounded. This procedure was repeated two
times in order to lower a possible impurity phase content. The obtained
sample was investigated by x-ray diffractometry to estimate its quality.
The diffractometer pattern (Fig.~\ref{xray Diffraktogramm}) showed
small impurity concentrations mainly resulting from MgO and  unreacted
carbon crystallized in form of graphite. The lattice constant of the
prepared sample was determined to be \( \textrm{a}=0.38107(1)\textrm{ nm} \)
using the Rietveld computer program FULLPROF.\cite{fullprof02} According
to Ref. \onlinecite{ren02} this indicates that the nearly single phase sample corresponds to the superconducting modification of \( \textrm{MgC}_{\textrm{x}}\textrm{Ni}_{3} \).

The superconducting transition of the sample was investigated by measurements
of the electrical resistance, the ac susceptibility and the specific
heat. For the electrical resistance measurement a piece cut from the
initially prepared pellet with 5mm in length and a cross section of
approximately 1mm\( ^{2} \) was measured in magnetic fields up to
16 T using the standard four probe method with current densities between
0.2 and 1 A/cm\( ^{2} \). The ac susceptibility and the specific
heat measurements were performed on other pieces from the same pellet
in magnetic fields up to 9 T.

\section{Results}

\subsection{Superconducting transition and upper critical field}

\begin{figure}[tbph]
{\centering \resizebox*{8cm}{6cm}{\includegraphics{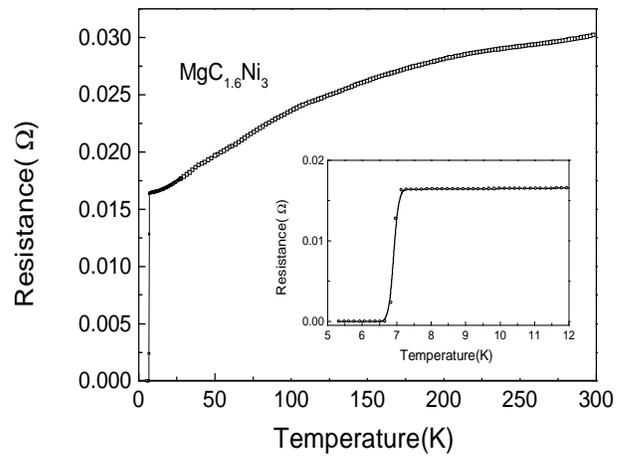}} \par}

\caption{Temperature dependence of resistance of MgC\protect\( _{1.6}\protect \)Ni\protect\( _{3}\protect \)
at zero applied magnetic field. The inset shows the superconducting transition region.\label{Bild - Gesamtwiderstand}}
\end{figure}
In Fig.~\ref{Bild - Gesamtwiderstand}, the temperature dependence
of the electrical resistance of the investigated sample is shown.
A superconducting transition with an onset (midpoint) value of \( T_{\textrm{c}}=7.0\textrm{ K} \)
(6.9 K) is observed (see inset of Fig.~\ref{Bild - Gesamtwiderstand})
which is consistent with the onset of the superconducting transition
at \( T_{\textrm{c}}=7.0\textrm{ K} \) determined from ac susceptibility.
It should be noted that the sample shown in Fig.~\ref{Bild - Gesamtwiderstand}
has a resistivity of \( \rho _{300\textrm{K}}=2.1\textrm{ m}\Omega \textrm{cm} \)
at 300 K which is much too large in order to be intrinsic. On the
other hand, its residual resistance ratio \( RRR=R(300\textrm{K})/R(8\textrm{K})=1.85 \)
and the shape of the \( R(T) \) curve are typical for MgCNi\( _{3} \)
samples.\cite{he01} A possible explanation for the high resistivity
of the investigated sample which was not subjected to high pressure
sintering is a relatively large resistance of the grain boundaries.

\begin{figure}[tbph]
{\centering \resizebox*{7cm}{6cm}{\rotatebox{-90}{\includegraphics{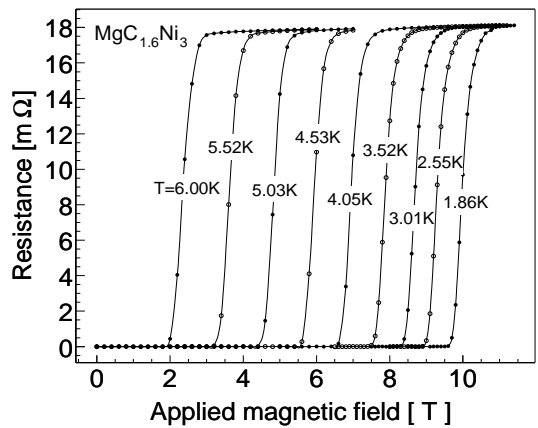}}} \par}

\caption{Field dependence of resistance of MgC\protect\( _{1.6}\protect \)Ni\protect\( _{3}\protect \)
measured at several temperatures. \label{Bild - Widerstand vs Magnetfeld}}
\end{figure}
\begin{figure}[tbph]
{\centering \resizebox*{8cm}{6.5cm}{\includegraphics{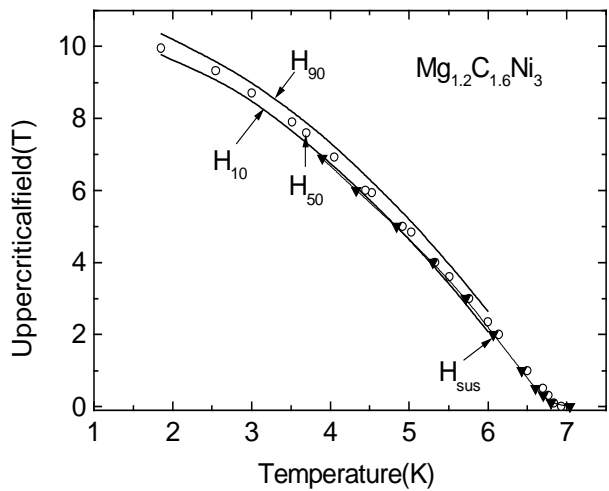}} \par}
\caption{Temperature dependence of the upper critical field for
MgC\protect\( _{1.6}\protect \)Ni\protect\( _{3}\protect \).  The
circles show the midpoint value \( H_{50} \) of the resistivity in the
normal state.  The two lines labelled \( H_{10} \) and \( H_{90} \)
denote 10\% and 90\% of the normal state resistivity. The triangles
represent the upper critical field determined from the onset of ac
susceptibility. \label{Bild - Widerstand & Suszeptibilit=E4t}}
\end{figure}

The field dependence of the electrical resistance of the same sample
is shown in Fig.~\ref{Bild - Widerstand vs Magnetfeld} for several
temperatures between \( 6.0 \) and \( 1.9\textrm{K} \). A sharp
transition is observed which remains almost unchanged down to low
temperatures.  In Fig.~\ref{Bild - Widerstand & Suszeptibilit=E4t},
the fields \( H_{10} \), \( H_{50} \) and \( H_{90} \) defined at
10\%, 50\% and 90\% of the normal-state resistance are plotted as
function of temperature. Identical results have been found from
resistance--vs.--temperature transition curves measured at different
magnetic fields. Additionally, Fig.~\ref{Bild - Widerstand &
Suszeptibilit=E4t} shows upper critical field data determined from ac
susceptibility measurements. The onset of superconductivity was used
to define \( H_{\textrm{c}2} \) from ac susceptibility.

It is clearly seen that for the investigated sample \(
H_{\textrm{c}2}^{\textrm{sus}} \) (\( H_{\textrm{c}2} \) obtained from
susceptibility) agrees approximately with \( H_{10} \). A similar
behavior was already observed for MgB\( _{2} \), whereas in the case
of rare--earth nickel borocarbides the onset of superconductivity
determined from ac susceptibility was typically found to agree well
with the midpoint value (\( H_{50} \)) of the normal state
resistivity. The width \( \Delta H=H_{90}-H_{10} \) of the
superconducting transition curves in Fig.~\ref{Bild - Widerstand vs
Magnetfeld} (and Fig.~\ref{Bild - Widerstand & Suszeptibilit=E4t})
remains, with \( \Delta H\simeq 0.6\textrm{ T} \), almost unchanged
down to low temperatures. A polycrystalline sample of a strongly
anisotropic superconductor shows a gradual broadening of the
superconducting transition with decreasing temperature as was
observed, for example, for MgB\( _{2} \).\cite{fuchs01} Therefore, the
nearly constant transition width \( \Delta H \) observed for the
investigated sample can be considered as an indication of a rather
small anisotropy of \( H_{c2}(T) \) in MgCNi\( _{3} \).

The extrapolation of \( H_{90}(T) \) to \( T=0 \) yields an upper
critical field of \( H_{\textrm{c}2}(0)\simeq 11.3\textrm{ T} \).  The
observed temperature dependence of the upper critical field is typical
for \( H_{\textrm{c}2}(T) \) data reported for MgCNi\( _{3} \) and was
described\cite{li01,lin,mao} within the standard WHH
model\cite{werthammer66} by conventional superconductivity in the
dirty limit. The WHH model predicts a relation \(
H_{\textrm{c}2}(0)\propto \left( - dH_{\textrm{c}2}/dT\right)
_{T=T_{\textrm{c}}}\cdot T_{\textrm{c}} \).  Available data for
MgCNi\( _{3} \), including those presented in this paper show a strong
variation of \( H_{\textrm{c}2}(0) \) with \( T_{\textrm{c}} \) (see
Fig.~5), whereas \( \left( dH_{\textrm{c}2}/dT\right)
_{\textrm{T}_{\textrm{c}}}\approx -(2.65\pm 0.2)\textrm{ T}/\textrm{K}
\) remains almost unchanged. Considering the data in Fig.~5, the
linear dependence of \( H_{\textrm{c}2}(0) \) on \( T_{\textrm{c}} \)
predicted for constant \( \left( dH_{\textrm{c}2}/dT\right)
_{T_{\textrm{c}}} \) by the WHH model can be ruled out.

\begin{figure}[tbph]
{\centering \resizebox*{8cm}{6.5cm}{\includegraphics{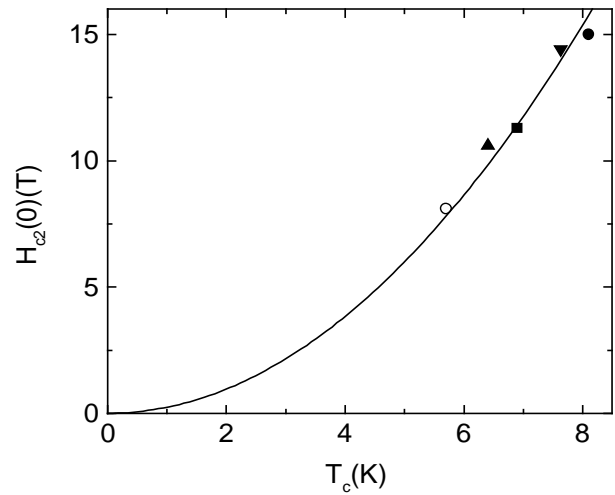}} \par}

\caption{Upper critical field at zero temperature vs. superconducting
transition temperature for the MgC\protect\( _{1.6}\protect
\)Ni\protect\( _{3}\protect \) sample of this work ( $\blacksquare$ )
and of several MgCNi\protect\( _{3}\protect \) samples reported in
Ref.~3 ( $\bigcirc$ ), Ref.~4 ( $\bullet$ ), Ref.~5 (
$\blacktriangledown$ ) and Ref.~6 ( $\blacktriangle$ ). The
experimental data can be described by a quadratic law (solid
line). \label{Bild - quadratisches Gesetz}}
\end{figure}
The solid line in Fig.~\ref{Bild - quadratisches Gesetz} corresponds
to a quadratic law \( H_{\textrm{c}2}(0)\sim T_{\textrm{c}}^{2} \)
which is the benchmark for the clean limit. Indeed in the isotropic
single band s--wave clean limit one has\cite{shulga}\begin{equation}
\label{s-Band Hc2}
H_{\textrm{c}2}(0)\left[ \textrm{Tesla}\right] =0.0237\frac{\left( 1+\lambda \right) ^{2.2}T_{\textrm{c}}^{2}\left[ \textrm{K}\right] }{v_{\textrm{F}}^{2}\left[ 10^{5}\frac{\textrm{m}}{\textrm{s}}\right] }
\end{equation}
Compared with WHH the effect of strong coupling (measured by the dimensionless
electron-phonon coupling constant \( \lambda  \)) is an enhancement
of \( H_{\textrm{c}2} \) primarily through the renormalization
of the bare Fermi velocity \( v_{\textrm{F}}\rightarrow v_{\textrm{F}}/(1+\lambda ) \)
and a weak enhancement factor coming from the energy dependence of
the gap function. The strong coupling (\( \lambda \approx 3.4 \)
near the Brillouin zone boundary and zero at the zone center) to the
rotational mode near 13 meV proposed in Ref.~\onlinecite{singh} gives an
upper limit for  a strong coupling required to reproduce upper
critical fields as high as \( 12\ldots 15\textrm{ T} \). Using the
averaged Fermi velocities of \( 1.45\cdot 10^{5}\textrm{ m}/\textrm{s} \)
for the rounded cube--like Fermi surface shown in Fig.~\ref{FSDOS} we arrive
at \( \lambda \approx 2.5\ldots 3 \).

\subsection{Specific heat}

\begin{figure}[tbph]
{\centering \resizebox*{8cm}{7cm}{\rotatebox{-90}{\includegraphics{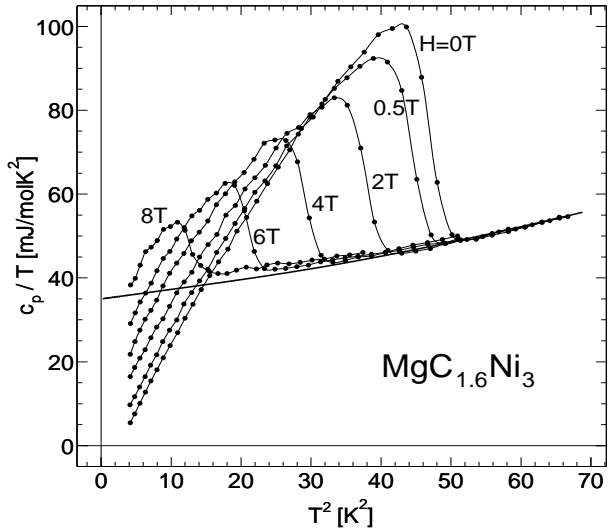}}} \par}

\caption{Specific heat data \( \textrm{c}_{\textrm{p}}/T \) vs. \( T^{2} \) of MgC\protect\( _{1.6}\protect \)Ni\protect\( _{3}\protect \)  measured at various magnetic fields up to \( 8\textrm{T} \). The
solid line is a fit of Eq.(\ref{spezifische Waerme cn}) to the data
for \protect\( H=0\protect \) above \protect\( T_{\textrm{c}}\protect \).
Its intersection with the ordinate gives the Sommerfeld parameter \protect\( \gamma _{\textrm{N}}\protect \).\label{Bild - spezifische W=E4rme komplett}}
\end{figure}
\begin{figure}[tbph]
{\centering \resizebox*{8cm}{6cm}{\includegraphics{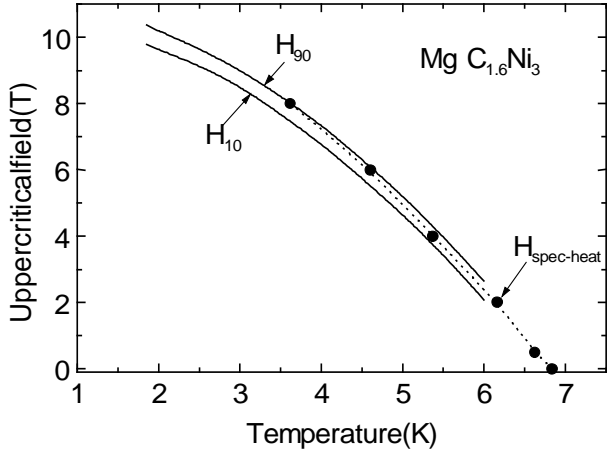}} \par}

\caption{Comparison of upper critical field data determined from
specific heat ( $\bullet$ ) and resistance measurements for
MgC\protect\( _{1.6}\protect \)Ni\protect\( _{3}\protect \).  \(
H_{10} \) and \( H_{90} \) were determined at 10\% and 90\% of the
normal state resistivity, respectively. An entropy conserving
construction was used to determine the upper critical field from the
specific heat data of Fig.~\ref{Bild - spezifische W=E4rme
komplett}. \label{Bild - Widerstand & spezifische W=E4rme}}
\end{figure}

Specific heat measurements were performed in order to get additional
information about \( H_{\textrm{c}2}(T) \), the superconducting pairing
symmetry and the strength of the electron--phonon coupling from thermodynamic
data. In Fig.~\ref{Bild - spezifische W=E4rme komplett} specific
heat data, \( \textrm{c}_{\textrm{p}}/T \) vs. \( T^{2} \), are
shown for applied magnetic fields up to \( 8\textrm{T} \). The upper
critical fields, \( \textrm{H}_{\textrm{c}2}(T) \), determined from
the specific heat data, are shown in Fig.~\ref{Bild - Widerstand & spezifische W=E4rme}.
It is clearly seen that the \( H_{c2}(T) \) data obtained from the
specific heat are located in the small field range between the \( H_{90}(T) \)
and \( H_{10}(T) \) curves.

\begin{figure*}[tbph]
{\centering
\resizebox*{16cm}{10cm}{\rotatebox{-90}{\includegraphics{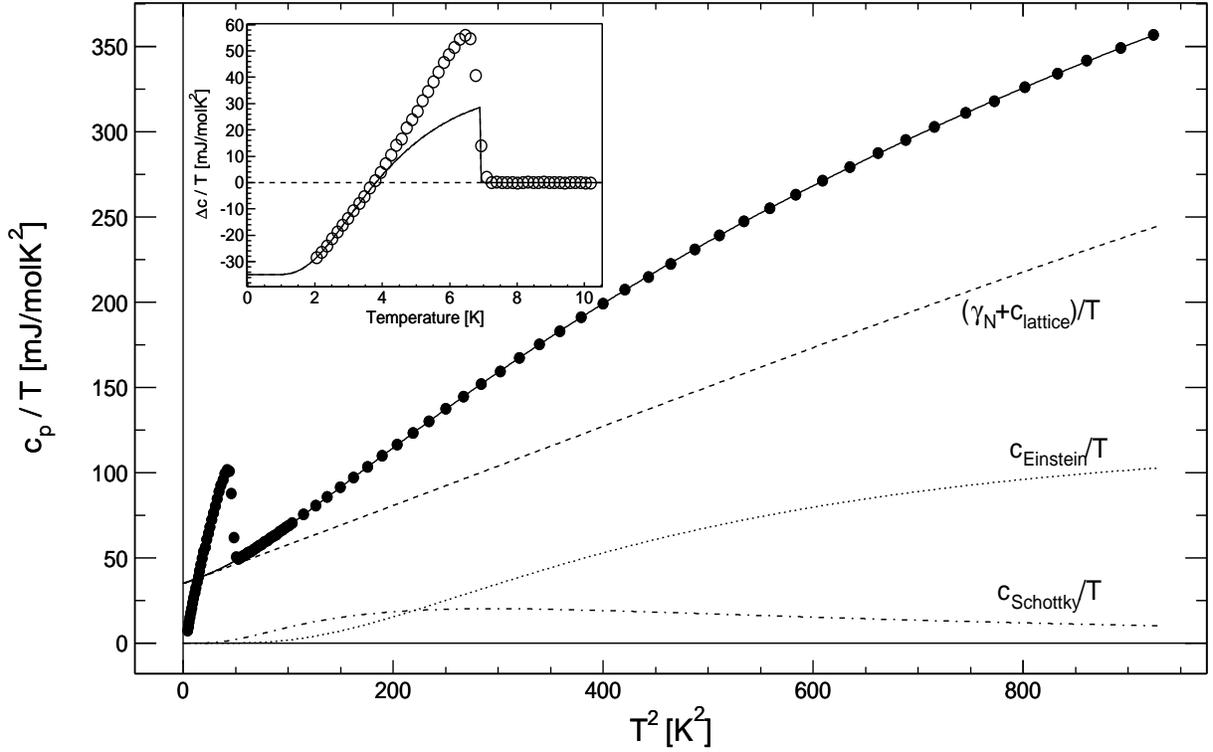}}}
\par}
\caption{Specific heat data \( \textrm{c}_{\textrm{p}}/T \) vs. \(
T^{2} \) of MgC\protect\( _{1.6}\protect \)Ni\protect\( _{3}\protect
\) at zero magnetic field. According to Eq.(\ref{spezifische Waerme
cn}), the dashed line represents the lattice contribution, the dotted
line shows the contribution from a rotational phonon mode (see text)
and the dash-dotted line gives a Schottky contribution. Inset:
Contribution of the superconducting electrons to the specific heat
\protect\( \Delta \textrm{c}/T=\left(
\textrm{c}_{\textrm{p}}-\textrm{c}_{\textrm{n}}\right) /T\protect \).
The solid line in the inset is a fit of the BCS expression to the data
with parameters according to Eq. (\ref{BCS Expression 2}). The
conservation of entropy was confirmed by integrating of \protect\(
\Delta \textrm{c}/T\protect \) in the temperature range \protect\(
0<T<T_{\textrm{c}}\protect \) (according to the data above \protect\(
2\textrm{K}\protect \) and the solid line below \protect\(
2\textrm{K}\protect \)).\label{Bild - spezifische W=E4rme H=3D0}}
\end{figure*}

In order to describe the experimental data of the normal state specific
heat in zero field between \( T_{\textrm{c}} \) and \( 30K \) (see
Fig.~\ref{Bild - spezifische W=E4rme H=3D0}) the expression\begin{equation}
\label{spezifische Waerme cn}
\textrm{c}_{\textrm{n}}(T)=\gamma _{\textrm{N}}T+\textrm{c}_{\textrm{lattice}}+\textrm{c}_{\textrm{Einstein}}+n\textrm{c}_{\textrm{Schottky}}
\end{equation}
was used. The linear--in--T term is due to the electronic contribution
with \( \gamma _{\textrm{N}} \) as the Sommerfeld parameter. The
second term, \( \textrm{c}_{\textrm{lattice}}=\beta T^{3}+\delta T^{5} \)
represents the phonon contribution. This extension of the usual Debye
approximation \( \textrm{c}_{\textrm{lattice}}\propto T^{3} \) includes
deviations from the linear dispersion of the acoustic modes and is
required in order to describe the phonon contribution to the specific
heat in an extended temperature range. The same procedure has already
been used to describe the specific heat of MgB\( _{2} \).\cite{waelti01}
An additional single Einstein mode, \( \textrm{c}_{\textrm{Einstein}} \),
was used to take into account the contribution of the previously mentioned
lowest energy optical mode near \( 13\textrm{meV} \).\cite{singh}
The last part, \( n\textrm{c}_{\textrm{Schottky}} \), represents the
contribution of a conventional two-level Schottky model which was
used to describe the slight upturn of the specific heat just above
the superconducting transition temperature \( T_{\textrm{c}} \) (see
Fig.~\ref{Bild - spezifische W=E4rme H=3D0}). This feature possibly
results from contributions of paramagnetic impurities like unreacted
Ni (the prefactor \( n \) gives the concentration of these impurities).

The fitting procedure returned a Sommerfeld parameter \( \gamma _{\textrm{N}}=35\textrm{mJ}/\textrm{molK}^{2} \)
which is close to the reported values of \( \gamma _{\textrm{N}}=27.6 \)
\cite{mao} and \( 33\textrm{mJ}/\textrm{molK}^{2} \).\cite{lin}
The Debye temperature was found to be \( \Theta _{\textrm{D}}=289\textrm{K} \)
and the energy of the optical phonon mode resulting from the fit is
\( 9.5\textrm{meV} \) which slightly deviates from the theoretically
proposed value (\( 13\textrm{meV}) \).\cite{singh} The paramagnetic
impurity concentration derived from the fit is \( n\approx 4.5\% \)
and the energy gap of the Schottky anomaly is \( 5.3\textrm{meV} \).
As shown in Fig.~\ref{Bild - spezifische W=E4rme H=3D0}, the experimental
data are very well described by Eq. (\ref{spezifische Waerme cn})
in the investigated temperature range \( T_{\textrm{c}}<T<30\textrm{K} \).
The jump \( \Delta \textrm{c} \) of the specific heat at \( T_{\textrm{c}} \)
(see inset of Fig.~\ref{Bild - spezifische W=E4rme H=3D0})
is given by the difference between the experimental data, \( \textrm{c}_{\textrm{p}} \)
and the normal specific heat contribution, \( \textrm{c}_{\textrm{n}} \).

The transition temperature \( T_{\textrm{c}}=6.83\textrm{K} \) 
obtained from the inset of  Fig.~\ref{Bild - spezifische W=E4rme H=3D0}
agrees approximately with the transition
temperatures \( T_{\textrm{c}}=7.0\textrm{K} \) and \(
T_{\textrm{c}}=6.9\textrm{K} \) derived from ac susceptibility and
from resistance data, respectively.  In the low temperature region the experimental data 
for \(\Delta \textrm{c}/T \) versus \( T \) can be described by the BCS--like expression
\begin{equation}
\label{BCS Expression 2}
\Delta \textrm{c}=7.95\gamma _{\textrm{N}}T_{\textrm{c}}\exp \left( -\frac{\Delta (0)}{k_{\textrm{B}}T}\right) -\gamma _{\textrm{N}}T
\end{equation}
using \( 2\Delta (0)/k_{\textrm{B}}T_{\textrm{c}}=2.96 \) instead
of \( 2\Delta _{\textrm{BCS}}(0)/k_{\textrm{B}}T_{\textrm{c}}=3.52 \)
predicted by the BCS model. Above \( T=4\textrm{K} \) the fit of
Eq. (\ref{BCS Expression 2}) starts to deviate from the experimental
curve resulting in a much lower value for the jump at \( T=T_{\textrm{c}} \)
than observed experimentally. Notice that the experimental value of
the jump, \( \Delta \textrm{c}/\gamma _{\textrm{N}}T_{\textrm{c}}=1.6 \), nearly corresponds
to the BCS value \( \Delta \textrm{c}/\gamma _{\textrm{N}}T_{\textrm{c}}=1.43 \) 
indicating weak electron--phonon coupling.
This contrasts to the strong coupling derived from Eq. (\ref{s-Band Hc2})
and \( H_{\textrm{c}2}(0) \) data. A natural explanation for this
discrepancy is the two--band character of MgCNi\( _{3} \) which will
be discussed in the next section.

\begin{figure}[tbph]
{\centering
\resizebox*{8cm}{5.33cm}{\rotatebox{-90}{\includegraphics{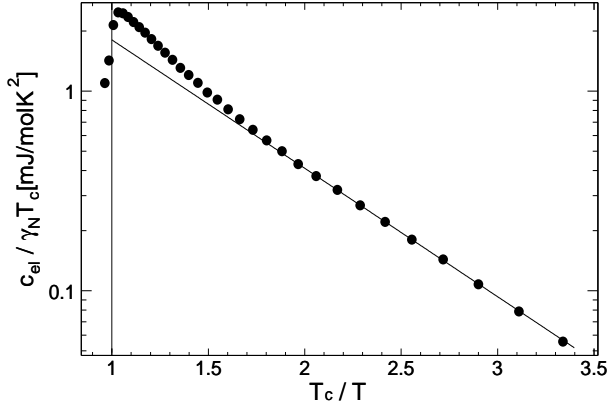}}}
\par}
\caption{Normalized electronic specific heat contribution of
MgC\protect\( _{1.6}\protect \)Ni\protect\( _{3}\protect \) vs. \(
T_{\textrm{c}}/T \). The comparison of the data with the BCS
expression (solid line) corresponding to Eq. (\ref{BCS Expression 2})
clearly shows the exponential temperature dependence of the electronic
specific heat at low temperatures. \label{Bild BCS - Thermodynamik}}
\end{figure}

To examine the temperature dependence of the electronic specific heat
\begin{equation}
\label{Formel c_el}
\textrm{c}_{\textrm{el}}(T)=\Delta \textrm{c}+\gamma _{\textrm{N}}T
\end{equation}
 at \( H=0 \) in detail, \( \textrm{c}_{\textrm{el}}(T)/\gamma
_{\textrm{N}}T_{\textrm{c}} \) is plotted logarithmically vs. \(
T_{\textrm{c}}/T \) in Fig.~\ref{Bild BCS - Thermodynamik}.  It is
clearly seen, that the experimental data at low temperatures (\(
T_{\textrm{c}}/T>2 \)) follow the modified BCS expression (Eq.
(\ref{BCS Expression 2}), solid line). We found that this exponential
law is not affected by the low temperature branch of the Schottky term
in Eq. (\ref{spezifische Waerme cn}) which has an exponential
temperature dependence, too. The exponential temperature behavior of
the electronic specific heat at low temperatures found for MgCNi\(
_{3} \) is a strong indication for s--wave superconductivity in this
compound.

\begin{figure}[tbph]
{\centering \resizebox*{8cm}{6.5cm}{\includegraphics{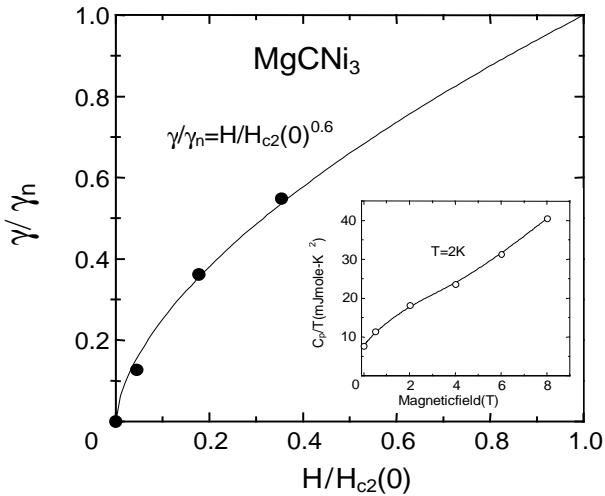}} \par}

\caption{Field dependence of the specific heat contribution
\protect\( \gamma (H) \) of the vortex core electrons in the mixed
state normalized by the Sommerfeld parameter \( \gamma _{\textrm{N}}
\) and \protect\(H_{\textrm{c}2}(0)\protect \). The black line is a
fit of \protect\( \gamma (H)/\gamma _{\textrm{N}}=\left(
H/H_{\textrm{c}2}(0)\right) ^{0.6}\protect \) to the experimental
data.  Inset: Field dependence of the specific heat \protect\(
\textrm{c}_{\textrm{p}}/T\protect \) at \protect\(
T=2\textrm{K}\protect \). \label{Bild - gamma von H}}
\end{figure}

In the superconducting state, a linear--in--T electronic specific
heat contribution \( \gamma (H)T \) arises from the normal conducting
cores of the flux lines for applied magnetic fields \( H>H_{\textrm{c}1} \).
This contribution can be expressed as \( \gamma (H)T=\textrm{c}_{\textrm{p}}(T,H)-\textrm{c}_{\textrm{p}}(T,0) \),\cite{sonier98}
where \( \textrm{c}_{\textrm{p}}(T,0) \) is the specific heat in
the Meissner state. Specific heat data for MgC\(_{1.6}\)Ni\( _{3} \) at \( T=2\textrm{K} \)
(see inset of Fig.~10) were analyzed in order to derive the field
dependence of \( \gamma (H) \). In Fig.~\ref{Bild - gamma von H},
the obtained \( \gamma (H)/\gamma _{\textrm{N}} \) is plotted against
\( H/H_{\textrm{c}2}(0) \) using the Sommerfeld parameter 
\( \gamma _{\textrm{N}}=35\textrm{mJ}/\textrm{molK}^{2} \)
and \( H_{\textrm{c}2}(0)=11.3\textrm{T} \). Not shown in this plot
are high-field data which are influenced by an additional contribution
to the specific heat in the normal state arising in magnetic fields.
This contribution which is clearly seen in Fig.~\ref{Bild - spezifische W=E4rme komplett}
as deviation of the \( \textrm{c}_{\textrm{p}}/T \) data from the
solid line causes a shift of the high-field \( \textrm{c}_{\textrm{p}}/T \)
data in the superconducting state to higher values. The origin of
this effect is not yet understood. Its influence on the field dependence
of \( \textrm{c}_{\textrm{p}}/T \) at \( T=2\textrm{K} \) is illustrated
in the inset of Fig.~\ref{Bild - gamma von H}. At low magnetic fields,
a negative curvature is observed which starts to change its sign at
fields above \( 4\textrm{T} \). 

The low-field data of \( \textrm{c}_{\textrm{p}}/T \) shown in
Fig.~\ref{Bild - gamma von H} can be described by the expression \(
\gamma /\gamma _{\textrm{N}}=\left( H/H_{\textrm{c}2}(0)\right) ^{0.6}
\) which differs from the linear \( \gamma (H) \) law expected for
isotropic s-wave superconductors in the dirty limit. A non-linear
field dependence close to \( \gamma (H)\propto H^{0.5} \) has been
reported for some unconventional superconductors with gap nodes in the
quasiparticle spectrum of the vortex state as in \(
\textrm{YBa}_{2}\textrm{Cu}_{3}\textrm{O}_{7} \),\cite{wright99} and
in the heavy fermion superconductor \( \textrm{UPt}_{3}
\),\cite{ramirez95} but also in some clean s-wave superconductors such
as \( \textrm{CeRu}_{2} \),\cite{hedo98} \( \textrm{NbSe}_{2}\)
\cite{sonier98,nohara99} and the borocarbides \(
\textrm{RNi}_{2}\textrm{B}_{2}\textrm{C } \) \(
(\textrm{R}=\textrm{Y},\textrm{ Lu}) \).\cite{nohara97,lipp02}
Delocalized quasiparticle states around the vortex cores, similar to
these in d-wave superconductors, seem to be responsible for the
non--linear \( \gamma (H) \) dependence in the
borocarbides.\cite{izawa01,boaknin01}

\section{Discussion}
\begin{figure}[tbph]
{\centering \resizebox*{8cm}{!}{\includegraphics{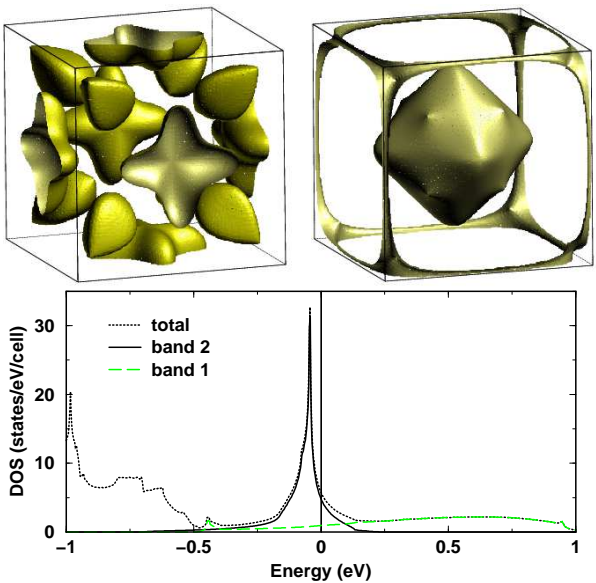}} \par}
\caption{The two Fermi surface sheets of MgCNi$_3$ and the
corresponding band resolved density of states near the Fermi
level. ``Band 1'' corresponds to the  Fermi surface sheet in the topright panel, ``band
2'' to the Fermi surface in the topleft panel.\label{FSDOS}}
\end{figure}

In principle, the upper critical field data found for MgCNi\( _{3} \)
could be understood both in the s- and p-wave scenario. The high
magnitude of the upper critical fields might be achieved in s-wave
superconductors, but also in a clean limit weak coupling p-wave
case employing the {}``slow'' holes on the {}``four-leafed
clover''--like Fermi surface sheets with \( 0.5\cdot
10^{5}\textrm{m}/\textrm{s} \) (see Fig.~\ref{FSDOS}). Following Maki
et al. \cite{maki01} an additional numerical factor 
\(1.3 \) should be introduced in this case in Eq.  (\ref{s-Band
Hc2}). The different temperature dependence of the electronic specific
heat at low temperatures in $s$- and $p$-wave superconductors allows
the discrimination between predominant s- and p-wave scenarios.  As
mentioned above, the exponential temperature behavior of the
electronic specific heat found for MgCNi\( _{3} \) is a strong
indication for $s$-wave superconductivity in this compound.

It is convenient to rewrite Eq. (\ref{s-Band Hc2}) using
experimentally accessible quantities as the plasma energy $\omega _{\textrm{pl}}$, the volume
of the unit cell $V$, and the Sommerfeld constant $\gamma_N$.  Then,
Eq. (\ref{s-Band Hc2 Formel Q-Check}) provides a
criterion for a superconductor which can be described in the clean
limit within the isotropic single band model:
\begin{equation}
\label{s-Band Hc2 Formel Q-Check}
Q=\frac{3.6\omega _{\textrm{pl}}^{2}\left[ \textrm{eV}^{2}\right] H_{\textrm{c}2}(0)\left[ \textrm{T}^{2}\right] V\left[ \left( 10^{-10}m\right) ^{3}\right] }{\gamma _{\textrm{N}}\left[ \textrm{mJ}/\textrm{molK}^{2}\right] T_{\textrm{c}}^{2}\left[ \textrm{K}\right] \left( 1+\lambda \right) ^{1.4}}
\end{equation}
If $Q$ differs significantly from 1,  a more complex
model should be considered.

Some values for \( Q \) are summarized in Tab. \ref{Tabelle - Q check}.

\begin{table}[tbph]
{\centering {\large \begin{tabular}{|c||c|c|c|c|c|c||c|}
\hline 
&
\( \omega _{\textrm{pl}} \)&
\( H_{\textrm{c}2}(0) \)&
\( V \)&
\( \gamma _{\textrm{N}} \)&
\( T_{\textrm{c}} \)&
\( \lambda  \)&
\( Q \)\\
\hline 
\( \textrm{Nb} \)&
\( 9.9 \)&
\( 0.35 \)&
\( 18 \)&
\( 7.8 \)&
\( 9.3 \)&
\( 0.9 \)&
\( 1.4 \)\\
\( \textrm{YNi}_{2}\textrm{B}_{2}\textrm{C} \)&
\( 4.0 \)&
\( 10 \)&
\( 64 \)&
\( 19 \)&
\( 15 \)&
\( 0.7 \)&
\( 4.1 \)\\
\( \textrm{MgB}_{2} \)&
\( 7.0 \)&
\( 17 \)&
\( 29 \)&
\( 3.0 \)&
\( 40 \)&
\( 0.8 \)&
\( 7.9 \)\\
\( \textrm{MgB}_{2} \)&
\( 7.0 \)&
\( 17 \)&
\( 29 \)&
\( 3.0 \)&
\( 40 \)&
\( 2.5 \)&
\( 3.1 \)\\
\( \textrm{MgCNi}_{3} \)&
\( 3.2 \)&
\( 15 \)&
\( 56 \)&
\( 35 \)&
\( 8 \)&
\( 1.0 \)&
\( 5.2 \)\\
\( \textrm{MgCNi}_{3} \)&
\( 3.2 \)&
\( 15 \)&
\( 56 \)&
\( 35 \)&
\( 8 \)&
\( 2.5 \)&
\( 2.4 \)\\
\hline
\end{tabular}}\large \par}

\caption{Values of \protect\( Q\protect \) and further parameters (see
text) for selected superconductors. \protect\( Q\protect \) is a
measure for the applicability of the isotropic single band model. A
deviation from \protect\( Q\approx 1\protect \) indicates the need for
a more complex model. For MgB$_2$ and MgCNi$_3$, two limiting values for
$\lambda$ are consiered for illustration.
\label{Tabelle - Q check}}
\end{table}

The crystal structure of MgCNi\( _{3} \) can be seen as a three dimensional
analogue of the layered borocarbides. Thus it is instructive to compare
the results with the electronic specific heat dependence of the borocarbides
and the boronitrides which were found to be isostructural to the borocarbides.
The electronic specific heat of La\( _{3} \)Ni\( _{2} \)B\( _{2} \)N\( _{3-\delta } \)
exhibits an exponential temperature dependence\cite{michor96} as does
MgCNi\( _{3} \), while that of YNi\( _{2} \)B\( _{2} \)C or LuNi\( _{2} \)B\( _{2} \)C
follows a power law of the type 
 \( \textrm{c}_{\textrm{el}}=3\gamma _{\textrm{N}}\left( \frac{\textrm{T}}{\textrm{T}_{\textrm{c}}}\right) ^{\textrm{a}} \)
with an exponent \( \textrm{a}\approx 3 \).\cite{michor95,hong94}

The electron-phonon coupling constant \( \lambda \) in MgCNi\( _{3} \)
averaged over all Fermi surface sheets can be estimated from the
relation\begin{equation}
\label{gammaN-gamma0-relation}
\gamma _{\textrm{N}}=\frac{\pi ^{2}}{3}k_{\textrm{B}}^{2}\left( 1+\lambda \right) N(E_{\textrm{F}})=\gamma _{0}\left( 1+\lambda \right) 
\end{equation}
using the experimental value of the Sommerfeld constant \( \gamma
_{\textrm{N}}=35\textrm{mJ}/\textrm{molK}^{2} \) and the density of
states (DOS) at the Fermi level \( N(E_{\textrm{F}}) \) calculated
within the local density approximation. From the calculated density of states and the corresponding bare
specific heat coefficient, \( \gamma
_{0}=11\textrm{mJ}/\textrm{molK}^{2} \) one obtains a large
electron-phonon coupling constant \( \lambda =2.2 \).

According to band structure calculations using the full-potential
nonorthogonal local-orbital minimum-basis scheme\cite{koepernik99}
within the local density approximation, the total DOS can be
decomposed into a roughly \( 15\% \) and a \( 85\% \) contribution
stemming from the fast and the slow sheets of the Fermi surface,
respectively. Then, the total coupling constant averaged over all
Fermi surface sheets reads
\begin{equation}
\label{totale Kopplungskonstante}
\lambda _{\textrm{tot}}=\lambda _{1}
\frac{\textrm{N}_{1}(0)}{\textrm{N}(0)}+\lambda _{2}
\frac{\textrm{N}_{2}(0)}{\textrm{N}(0)}
\end{equation}
and with the aid of \( \lambda_{\textrm{tot}} =2.2 \) and \( \lambda _{1}\approx
2.5\ldots 3 \) estimated from the upper critical field \(
H_{\textrm{c}2}(0)\approx 12\ldots 15\textrm{T} \), respectively
(Eq. (\ref{s-Band Hc2})) one arrives at \( \lambda _{2}\approx
2.0\ldots 2.1 \).  The question then arises about the origin of these
strong mass enhancements in both bands. Obviously $\lambda_2$ cannot
be of electron-phonon nature alone, since otherwise one should ask,
why are $T_c$, the above mentioned gap, and $H_{c2}(0)$ so low?  So we
are forced to assume that $\lambda_2$ should be decomposed into an
electron-phonon contribution and a pair-breaking electron-paramagnon
$\lambda_{sf}$ (electron-spin fluctuation) one.  We adopt for the
electron-phonon part a typical transition metal value, say
$\lambda_{2ph} \sim$ 1.  From Fig.\ \ref{FSDOS}, a band
width of about 0.6 eV can be estimated for band~2. Taking the average of the
electron and the hole Fermi energies as a representative effective
Fermi energy of band 2 $\hbar \omega_{2F}\approx $ 0.3 eV can be
estimated. Then, according to the Berk-Schrieffer theory the
paramagnon spectral density should exhibit a maximum near
$\hbar\omega_{sf}\approx \hbar \omega_{2F} /S \sim$ 80 meV, where $S
\approx $ 4 is the Stoner factor.  Since this frequency exceeds
considerably the typical phonon frequencies, the effect of the
paramagnon pair breaking is somewhat reduced due to the
pseudopotential effect\cite{zhernov85}, well-known for the large bare
Coulomb repulsion $\mu$ where it produces the smaller pseudopotential
$\mu^*$-value due to the large logarithmic factor of about 5 entering
its denominator.  In the present case, this factor is reduced to 1.4,
only. Then the ``bare'' paramagnon coupling constant $\lambda_{sf}\sim
1$ is reduced to about $\lambda_{sf}^*\sim$ 0.4. Together with
standard Coulomb pseudopotential, e.g.\ 0.13, one nevertheless arrives
at a sizable suppressed pairing in band 2 which is responsible for the
observed small gap seen in the low $T$ specific heat data. The missing
od only very weakly pronounced
curvature in $H_{c2}(T)$ near $T_c$ points to weak interband coupling,
a posteriori justifying our one band estimate for $H_{c2}(0)$ given
above. A more quantitative study must await more detailed knowledge on
the phonon and paramagnon spectra.

\section{Conclusion}
The electronic specific heat shows an exponential temperature dependence at low temperatures which is a strong indication for s-wave pairing symmetry in this compound. Strong electron-phonon coupling has been derived from specific heat data and the calculated densitiy of states averaged over all Fermi surface sheets.
Comparing MgCNi$_3$ with other multi-band superconductors such as
MgB$_2$ or transition metal borocarbides, one concludes that the
disjoint Fermi surface sheets differ from each other not only in the
strength of the electron-phonon interaction or the Fermi velocities but to the
best of our knowledge also in strength of the depairing
interaction. Approaching the remarkable peak in the DOS slightly below
the Fermi energy, an interesting enhancement of the paramagnon
contribution followed by a gapless Fermi surface sheet and a possible $p$-wave
superconductivity mediated by ferromagnetic spin fluctuations might be
expected. In any case, the competing interplay of strong electron-phonon and
electron-paramagnon interactions together with a strongly energy dependent
DOS on disjoint, almost decoupled Fermi surface sheets is a great
challenge for future theoretical and experimental work.

\begin{acknowledgments}
The Sonderforschungsbereich 463 at the TU Dresden, the DAAD (H.R.),
the NSF (DMR-0114818) and the DFG are gratefully acknowledged for
financial support.  We thank T.\ Mishonov for discussions.

\end{acknowledgments}

\begin{thebibliography}{27}
\expandafter\ifx\csname natexlab\endcsname\relax\def\natexlab#1{#1}\fi
\expandafter\ifx\csname bibnamefont\endcsname\relax
  \def\bibnamefont#1{#1}\fi
\expandafter\ifx\csname bibfnamefont\endcsname\relax
  \def\bibfnamefont#1{#1}\fi
\expandafter\ifx\csname citenamefont\endcsname\relax
  \def\citenamefont#1{#1}\fi
\expandafter\ifx\csname url\endcsname\relax
  \def\url#1{\texttt{#1}}\fi
\expandafter\ifx\csname urlprefix\endcsname\relax\def\urlprefix{URL }\fi
\providecommand{\bibinfo}[2]{#2}
\providecommand{\eprint}[2][]{\url{#2}}

\bibitem[{\citenamefont{He et~al.}(2001)\citenamefont{He, Huang, Ramirez, Wang,
  Regan, Rogado, Hayward, Haas, Slusky, Inumara et~al.}}]{he01}
\bibinfo{author}{\bibfnamefont{T.}~\bibnamefont{He}},
  \bibinfo{author}{\bibfnamefont{Q.}~\bibnamefont{Huang}},
  \bibinfo{author}{\bibfnamefont{A.~P.} \bibnamefont{Ramirez}},
  \bibinfo{author}{\bibfnamefont{Y.}~\bibnamefont{Wang}},
  \bibinfo{author}{\bibfnamefont{K.~A.} \bibnamefont{Regan}},
  \bibinfo{author}{\bibfnamefont{N.}~\bibnamefont{Rogado}},
  \bibinfo{author}{\bibfnamefont{M.~A.} \bibnamefont{Hayward}},
  \bibinfo{author}{\bibfnamefont{M.~K.} \bibnamefont{Haas}},
  \bibinfo{author}{\bibfnamefont{J.~S.} \bibnamefont{Slusky}},
  \bibinfo{author}{\bibfnamefont{K.}~\bibnamefont{Inumara}},
  \bibnamefont{et~al.}, \bibinfo{journal}{Nature}
  \textbf{\bibinfo{volume}{411}}, \bibinfo{pages}{54} (\bibinfo{year}{2001}).

\bibitem[{\citenamefont{Rosner et~al.}(2002)\citenamefont{Rosner, Weht,
  Johannes, Pickett, and Tosatti}}]{rosner02}
\bibinfo{author}{\bibfnamefont{H.}~\bibnamefont{Rosner}},
  \bibinfo{author}{\bibfnamefont{R.}~\bibnamefont{Weht}},
  \bibinfo{author}{\bibfnamefont{M.~D.} \bibnamefont{Johannes}},
  \bibinfo{author}{\bibfnamefont{W.~E.} \bibnamefont{Pickett}},
  \bibnamefont{and} \bibinfo{author}{\bibfnamefont{E.}~\bibnamefont{Tosatti}},
  \bibinfo{journal}{Phys. Rev. Lett.} \textbf{\bibinfo{volume}{88}},
  \bibinfo{pages}{027001} (\bibinfo{year}{2002}).


\bibitem[{\citenamefont{Singh and Mazin}()}]{singh}
\bibinfo{author}{\bibfnamefont{D.~J.} \bibnamefont{Singh}} \bibnamefont{and}
  \bibinfo{author}{\bibfnamefont{I.}~\bibnamefont{Mazin}},
  \eprint{cond-mat/0105577}.

\bibitem[{\citenamefont{Dugdale and Jarlborg}(2001)}]{jarlborg}
\bibinfo{author}{\bibfnamefont{S.~B.} \bibnamefont{Dugdale}} \bibnamefont{and}
\bibinfo{author}{\bibfnamefont{T.} \bibnamefont{Jarlborg}},
  \bibinfo{journal}{Phys. Rev. B } \textbf{\bibinfo{volume}{64}},
  \bibinfo{pages}{100508} (\bibinfo{year}{2001}).


\bibitem[{\citenamefont{Li et~al.}()}]{li}
\bibinfo{author}{\bibfnamefont{S.~Y.} \bibnamefont{Li}} \bibnamefont{et~al.},
  \eprint{cond-mat/0104554}.

\bibitem[{\citenamefont{Li et~al.}(2001)\citenamefont{Li, Fan, Chen, Wang, Mo,
  Ruan, Xiong, Luo, Zhang, Li et~al.}}]{li01}
\bibinfo{author}{\bibfnamefont{S.~Y.} \bibnamefont{Li}},
  \bibinfo{author}{\bibfnamefont{R.}~\bibnamefont{Fan}},
  \bibinfo{author}{\bibfnamefont{X.~H.} \bibnamefont{Chen}},
  \bibinfo{author}{\bibfnamefont{C.~H.} \bibnamefont{Wang}},
  \bibinfo{author}{\bibfnamefont{W.~Q.} \bibnamefont{Mo}},
  \bibinfo{author}{\bibfnamefont{K.~Q.} \bibnamefont{Ruan}},
  \bibinfo{author}{\bibfnamefont{Y.~M.} \bibnamefont{Xiong}},
  \bibinfo{author}{\bibfnamefont{X.~G.} \bibnamefont{Luo}},
  \bibinfo{author}{\bibfnamefont{H.~T.} \bibnamefont{Zhang}},
  \bibinfo{author}{\bibfnamefont{L.}~\bibnamefont{Li}}, \bibnamefont{et~al.},
  \bibinfo{journal}{Phys. Rev. B} \textbf{\bibinfo{volume}{64}},
  \bibinfo{pages}{132505} (\bibinfo{year}{2001}).

\bibitem[{\citenamefont{Mao et~al.}()\citenamefont{Mao, Rosario, Nelson, Wu,
  Deac, Schiffer, and Liu}}]{mao}
\bibinfo{author}{\bibfnamefont{Z.~Q.} \bibnamefont{Mao}},
  \bibinfo{author}{\bibfnamefont{M.~M.} \bibnamefont{Rosario}},
  \bibinfo{author}{\bibfnamefont{K.~D.} \bibnamefont{Nelson}},
  \bibinfo{author}{\bibfnamefont{K.}~\bibnamefont{Wu}},
  \bibinfo{author}{\bibfnamefont{I.~G.} \bibnamefont{Deac}},
  \bibinfo{author}{\bibfnamefont{P.}~\bibnamefont{Schiffer}}, \bibnamefont{and}
  \bibinfo{author}{\bibfnamefont{Y.}~\bibnamefont{Liu}},
  \eprint{cond-mat/0105280(v3)}.

\bibitem[{\citenamefont{Lin et~al.}()\citenamefont{Lin, Ho, Huang, Lin, Zhang,
  Yu, Jin, and Yang}}]{lin}
\bibinfo{author}{\bibfnamefont{J.-Y.} \bibnamefont{Lin}},
  \bibinfo{author}{\bibfnamefont{P.~L.} \bibnamefont{Ho}},
  \bibinfo{author}{\bibfnamefont{H.~L.} \bibnamefont{Huang}},
  \bibinfo{author}{\bibfnamefont{P.~H.} \bibnamefont{Lin}},
  \bibinfo{author}{\bibfnamefont{Y.-L.} \bibnamefont{Zhang}},
  \bibinfo{author}{\bibfnamefont{R.-C.} \bibnamefont{Yu}},
  \bibinfo{author}{\bibfnamefont{C.-Q.} \bibnamefont{Jin}}, \bibnamefont{and}
  \bibinfo{author}{\bibfnamefont{H.~D.} \bibnamefont{Yang}},
  \eprint{cond-mat/0202034}.

\bibitem[{\citenamefont{Singer et~al.}(2001)\citenamefont{Singer, Imai, He,
  Hayward, and Cava}}]{singer01}
\bibinfo{author}{\bibfnamefont{P.~M.} \bibnamefont{Singer}},
  \bibinfo{author}{\bibfnamefont{T.}~\bibnamefont{Imai}},
  \bibinfo{author}{\bibfnamefont{T.}~\bibnamefont{He}},
  \bibinfo{author}{\bibfnamefont{M.~A.} \bibnamefont{Hayward}},
  \bibnamefont{and} \bibinfo{author}{\bibfnamefont{R.~J.} \bibnamefont{Cava}},
  \bibinfo{journal}{Phys. Rev. Lett.} \textbf{\bibinfo{volume}{87}},
  \bibinfo{pages}{257601} (\bibinfo{year}{2001}).

\bibitem[{ful()}]{fullprof02}
\eprint{Rietveld refinement program FULLPROF 2k}.

\bibitem[{\citenamefont{Ren et~al.}(2002)\citenamefont{Ren, Che, Jia, Chen, Ni,
  Liu, and Zhao}}]{ren02}
\bibinfo{author}{\bibfnamefont{Z.~A.} \bibnamefont{Ren}},
  \bibinfo{author}{\bibfnamefont{G.~C.} \bibnamefont{Che}},
  \bibinfo{author}{\bibfnamefont{S.~L.} \bibnamefont{Jia}},
  \bibinfo{author}{\bibfnamefont{H.}~\bibnamefont{Chen}},
  \bibinfo{author}{\bibfnamefont{Y.~M.} \bibnamefont{Ni}},
  \bibinfo{author}{\bibfnamefont{G.~D.} \bibnamefont{Liu}}, \bibnamefont{and}
  \bibinfo{author}{\bibfnamefont{Z.~X.} \bibnamefont{Zhao}},
  \bibinfo{journal}{Physica C} \textbf{\bibinfo{volume}{371}},
  \bibinfo{pages}{1} (\bibinfo{year}{2002}).

\bibitem[{\citenamefont{Fuchs et~al.}(2001)\citenamefont{Fuchs, Müller,
  Handstein, Nenkov, Narozhnyi, Eckert, Wolf, and Schultz}}]{fuchs01}
\bibinfo{author}{\bibfnamefont{G.}~\bibnamefont{Fuchs}},
  \bibinfo{author}{\bibfnamefont{K.-H.} \bibnamefont{Müller}},
  \bibinfo{author}{\bibfnamefont{A.}~\bibnamefont{Handstein}},
  \bibinfo{author}{\bibfnamefont{K.}~\bibnamefont{Nenkov}},
  \bibinfo{author}{\bibfnamefont{V.~N.} \bibnamefont{Narozhnyi}},
  \bibinfo{author}{\bibfnamefont{D.}~\bibnamefont{Eckert}},
  \bibinfo{author}{\bibfnamefont{M.}~\bibnamefont{Wolf}}, \bibnamefont{and}
  \bibinfo{author}{\bibfnamefont{L.}~\bibnamefont{Schultz}},
  \bibinfo{journal}{Sol. State Comm.} \textbf{\bibinfo{volume}{118}},
  \bibinfo{pages}{497} (\bibinfo{year}{2001}).

\bibitem[{\citenamefont{Werthammer et~al.}(1966)\citenamefont{Werthammer,
  Helfand, and Hohenberg}}]{werthammer66}
\bibinfo{author}{\bibfnamefont{N.~R.} \bibnamefont{Werthammer}},
  \bibinfo{author}{\bibfnamefont{E.}~\bibnamefont{Helfand}}, \bibnamefont{and}
  \bibinfo{author}{\bibfnamefont{P.~C.} \bibnamefont{Hohenberg}},
  \bibinfo{journal}{Phys. Rev.} \textbf{\bibinfo{volume}{147}},
  \bibinfo{pages}{295} (\bibinfo{year}{1966}).

\bibitem[{\citenamefont{Shulga and Drechsler}()}]{shulga}
\bibinfo{author}{\bibfnamefont{S.~V.} \bibnamefont{Shulga}} \bibnamefont{and}
  \bibinfo{author}{\bibfnamefont{S.-L.} \bibnamefont{Drechsler}},
  \eprint{cond-mat/0202172. A linear fit yields a negative interception in
  contradiction with the expected behavior in the dirty limit case. If one
  would adopt nevertheless a dirty limit scenario, the measured (reported in
  Ref. Cava residual resistivity of $\rho$(8K) = 1.14 m$\Omega$cm (40 m$\Omega$cm)
  corresponds to an upper critical field of the order of about 5000 (250)
  Tesla, where a plasma frequency of 3.25 eV 11 and our calculated averaged
  Fermi velocity of 1.45$\cdot10^{5}$ m/s have been adopted. Thus we conclude
  that the reported so far resistivity data do not reflect the intrinsic
  properties of polycrystalline grains.}

\bibitem[{\citenamefont{Wälti et~al.}(2001)\citenamefont{Wälti, Felder, Degen,
  Wigger, Monnier, Delley, and Ott}}]{waelti01}
\bibinfo{author}{\bibfnamefont{C.}~\bibnamefont{Wälti}},
  \bibinfo{author}{\bibfnamefont{E.}~\bibnamefont{Felder}},
  \bibinfo{author}{\bibfnamefont{C.}~\bibnamefont{Degen}},
  \bibinfo{author}{\bibfnamefont{G.}~\bibnamefont{Wigger}},
  \bibinfo{author}{\bibfnamefont{R.}~\bibnamefont{Monnier}},
  \bibinfo{author}{\bibfnamefont{B.}~\bibnamefont{Delley}}, \bibnamefont{and}
  \bibinfo{author}{\bibfnamefont{H.~R.} \bibnamefont{Ott}},
  \bibinfo{journal}{Phys. Rev. B} \textbf{\bibinfo{volume}{01}},
  \bibinfo{pages}{172515} (\bibinfo{year}{2001}).

\bibitem[{\citenamefont{Sonier et~al.}(1998)\citenamefont{Sonier, Hundley,
  Thompson, and Brill}}]{sonier98}
\bibinfo{author}{\bibfnamefont{J.~E.} \bibnamefont{Sonier}},
  \bibinfo{author}{\bibfnamefont{M.~F.} \bibnamefont{Hundley}},
  \bibinfo{author}{\bibfnamefont{J.~D.} \bibnamefont{Thompson}},
  \bibnamefont{and} \bibinfo{author}{\bibfnamefont{J.~W.} \bibnamefont{Brill}},
  \bibinfo{journal}{Phys. Rev. Lett.} \textbf{\bibinfo{volume}{82}},
  \bibinfo{pages}{4914} (\bibinfo{year}{1998}).

\bibitem[{\citenamefont{Wright et~al.}(1999)\citenamefont{Wright, Emerson,
  Woodfield, Gordon, Fisher, and Phillips}}]{wright99}
\bibinfo{author}{\bibfnamefont{D.~A.} \bibnamefont{Wright}},
  \bibinfo{author}{\bibfnamefont{J.~P.} \bibnamefont{Emerson}},
  \bibinfo{author}{\bibfnamefont{B.~F.} \bibnamefont{Woodfield}},
  \bibinfo{author}{\bibfnamefont{J.~E.} \bibnamefont{Gordon}},
  \bibinfo{author}{\bibfnamefont{R.~A.} \bibnamefont{Fisher}},
  \bibnamefont{and} \bibinfo{author}{\bibfnamefont{N.~E.}
  \bibnamefont{Phillips}}, \bibinfo{journal}{Phys. Rev. Lett.}
  \textbf{\bibinfo{volume}{82}}, \bibinfo{pages}{1550} (\bibinfo{year}{1999}).

\bibitem[{\citenamefont{Ramirez et~al.}(1995)\citenamefont{Ramirez, Stücheli,
  and Bucher}}]{ramirez95}
\bibinfo{author}{\bibfnamefont{A.~P.} \bibnamefont{Ramirez}},
  \bibinfo{author}{\bibfnamefont{N.}~\bibnamefont{Stücheli}}, \bibnamefont{and}
  \bibinfo{author}{\bibfnamefont{E.}~\bibnamefont{Bucher}},
  \bibinfo{journal}{Phys. Rev. Lett.} \textbf{\bibinfo{volume}{74}},
  \bibinfo{pages}{1218} (\bibinfo{year}{1995}).

\bibitem[{\citenamefont{Hedo}(1998)}]{hedo98}
\bibinfo{author}{\bibfnamefont{M.}~\bibnamefont{Hedo}}, \bibinfo{journal}{J.
  Phys. Soc. Japan} \textbf{\bibinfo{volume}{67}}, \bibinfo{pages}{272}
  (\bibinfo{year}{1998}).

\bibitem[{\citenamefont{Nohara et~al.}(1999)\citenamefont{Nohara, Sakai, and
  Takagi}}]{nohara99}
\bibinfo{author}{\bibfnamefont{M.}~\bibnamefont{Nohara}},
  \bibinfo{author}{\bibfnamefont{M.~I.~F.} \bibnamefont{Sakai}},
  \bibnamefont{and} \bibinfo{author}{\bibfnamefont{H.}~\bibnamefont{Takagi}},
  \bibinfo{journal}{J. Phys. Soc. Japan} \textbf{\bibinfo{volume}{68}},
  \bibinfo{pages}{1078} (\bibinfo{year}{1999}).

\bibitem[{\citenamefont{Nohara et~al.}(1997)\citenamefont{Nohara, Isshiki,
  Takagi, and Cava}}]{nohara97}
\bibinfo{author}{\bibfnamefont{M.}~\bibnamefont{Nohara}},
  \bibinfo{author}{\bibfnamefont{M.}~\bibnamefont{Isshiki}},
  \bibinfo{author}{\bibfnamefont{H.}~\bibnamefont{Takagi}}, \bibnamefont{and}
  \bibinfo{author}{\bibfnamefont{R.~J.} \bibnamefont{Cava}},
  \bibinfo{journal}{J. Phys. Soc. Japan} \textbf{\bibinfo{volume}{66}},
  \bibinfo{pages}{1888} (\bibinfo{year}{1997}).

\bibitem[{\citenamefont{Lipp et~al.}(2002)\citenamefont{Lipp, Schneider,
  Gladun, Drechsler, Freudenberger, Fuchs, Nenkov, Müller, Cichorek, and
  Gegenwart}}]{lipp02}
\bibinfo{author}{\bibfnamefont{D.}~\bibnamefont{Lipp}},
  \bibinfo{author}{\bibfnamefont{M.}~\bibnamefont{Schneider}},
  \bibinfo{author}{\bibfnamefont{A.}~\bibnamefont{Gladun}},
  \bibinfo{author}{\bibfnamefont{S.-L.} \bibnamefont{Drechsler}},
  \bibinfo{author}{\bibfnamefont{J.}~\bibnamefont{Freudenberger}},
  \bibinfo{author}{\bibfnamefont{G.}~\bibnamefont{Fuchs}},
  \bibinfo{author}{\bibfnamefont{K.}~\bibnamefont{Nenkov}},
  \bibinfo{author}{\bibfnamefont{K.-H.} \bibnamefont{Müller}},
  \bibinfo{author}{\bibfnamefont{T.}~\bibnamefont{Cichorek}}, \bibnamefont{and}
  \bibinfo{author}{\bibfnamefont{P.}~\bibnamefont{Gegenwart}},
  \bibinfo{journal}{Europhys.\ Lett.} \textbf{\bibinfo{volume}{58}},
  \bibinfo{pages}{435} (\bibinfo{year}{2002}).

\bibitem[{\citenamefont{Izawa et~al.}(2001)\citenamefont{Izawa, Shibata,
  Matsuda, Kato, Takeya, Hirata, van~der Beek, and Konczykowski}}]{izawa01}
\bibinfo{author}{\bibfnamefont{K.}~\bibnamefont{Izawa}},
  \bibinfo{author}{\bibfnamefont{A.}~\bibnamefont{Shibata}},
  \bibinfo{author}{\bibfnamefont{Y.}~\bibnamefont{Matsuda}},
  \bibinfo{author}{\bibfnamefont{Y.}~\bibnamefont{Kato}},
  \bibinfo{author}{\bibfnamefont{H.}~\bibnamefont{Takeya}},
  \bibinfo{author}{\bibfnamefont{K.}~\bibnamefont{Hirata}},
  \bibinfo{author}{\bibfnamefont{C.~J.} \bibnamefont{van~der Beek}},
  \bibnamefont{and}
  \bibinfo{author}{\bibfnamefont{M.}~\bibnamefont{Konczykowski}},
  \bibinfo{journal}{Phys. Rev. Lett.} \textbf{\bibinfo{volume}{86}},
  \bibinfo{pages}{1327} (\bibinfo{year}{2001}).

\bibitem[{\citenamefont{Boaknin et~al.}(2001)\citenamefont{Boaknin, Hill,
  Proust, Lupien, Taillefer, and Canfield}}]{boaknin01}
\bibinfo{author}{\bibfnamefont{E.}~\bibnamefont{Boaknin}},
  \bibinfo{author}{\bibfnamefont{R.~W.} \bibnamefont{Hill}},
  \bibinfo{author}{\bibfnamefont{C.}~\bibnamefont{Proust}},
  \bibinfo{author}{\bibfnamefont{C.}~\bibnamefont{Lupien}},
  \bibinfo{author}{\bibfnamefont{L.}~\bibnamefont{Taillefer}},
  \bibnamefont{and} \bibinfo{author}{\bibfnamefont{P.~C.}
  \bibnamefont{Canfield}}, \bibinfo{journal}{Phys. Rev. Lett.}
  \textbf{\bibinfo{volume}{87}}, \bibinfo{pages}{237001}
  (\bibinfo{year}{2001}).

\bibitem[{\citenamefont{Maki et~al.}(2001)\citenamefont{Maki, Puchkarev, and
  Wang}}]{maki01}
\bibinfo{author}{\bibfnamefont{K.}~\bibnamefont{Maki}},
  \bibinfo{author}{\bibfnamefont{E.}~\bibnamefont{Puchkarev}},
  \bibnamefont{and} \bibinfo{author}{\bibfnamefont{G.~F.} \bibnamefont{Wang}},
  \emph{\bibinfo{title}{High-T$_c$ superconductors and related Materials}},
  Series 3 High Technology v. 86, (Ed. S.-L. Drechsler and T. Mishonov),
  (\bibinfo{publisher}{Kluwer Academic Publisher Dordrecht},
  \bibinfo{year}{2001}).

\bibitem[{\citenamefont{Michor et~al.}(1996)\citenamefont{Michor,
  Krendelsberger, Hilscher, Bauer, C.Dusek, Hauser, Naber, Werner, Rogl, and
  Zanbergen}}]{michor96}
\bibinfo{author}{\bibfnamefont{H.}~\bibnamefont{Michor}},
  \bibinfo{author}{\bibfnamefont{R.}~\bibnamefont{Krendelsberger}},
  \bibinfo{author}{\bibfnamefont{G.}~\bibnamefont{Hilscher}},
  \bibinfo{author}{\bibfnamefont{E.}~\bibnamefont{Bauer}},
  \bibinfo{author}{\bibnamefont{C.Dusek}},
  \bibinfo{author}{\bibfnamefont{R.}~\bibnamefont{Hauser}},
  \bibinfo{author}{\bibfnamefont{L.}~\bibnamefont{Naber}},
  \bibinfo{author}{\bibfnamefont{D.}~\bibnamefont{Werner}},
  \bibinfo{author}{\bibfnamefont{P.}~\bibnamefont{Rogl}}, \bibnamefont{and}
  \bibinfo{author}{\bibfnamefont{H.~W.} \bibnamefont{Zanbergen}},
  \bibinfo{journal}{Phys. Rev. B} \textbf{\bibinfo{volume}{54}},
  \bibinfo{pages}{9408} (\bibinfo{year}{1996}).

\bibitem[{\citenamefont{Michor et~al.}(1995)\citenamefont{Michor, Holubar,
  Dusek, and Hilscher}}]{michor95}
\bibinfo{author}{\bibfnamefont{H.}~\bibnamefont{Michor}},
  \bibinfo{author}{\bibfnamefont{T.}~\bibnamefont{Holubar}},
  \bibinfo{author}{\bibfnamefont{C.}~\bibnamefont{Dusek}}, \bibnamefont{and}
  \bibinfo{author}{\bibfnamefont{G.}~\bibnamefont{Hilscher}},
  \bibinfo{journal}{Phys. Rev. B} \textbf{\bibinfo{volume}{52}},
  \bibinfo{pages}{16165} (\bibinfo{year}{1995}).

\bibitem[{\citenamefont{Hong et~al.}(1994)}]{hong94}
\bibinfo{author}{\bibfnamefont{N.~M.} \bibnamefont{Hong}} \bibnamefont{et~al.},
  \bibinfo{journal}{Physica C} \textbf{\bibinfo{volume}{227}},
  \bibinfo{pages}{85} (\bibinfo{year}{1994}).

\bibitem[{\citenamefont{Koepernik and H. Eschrig}(1999)}]{koepernik99}
\bibinfo{author}{\bibfnamefont{K} \bibnamefont{Koepernik}} \bibnamefont{and}
\bibinfo{author}{\bibfnamefont{H.} \bibnamefont{Eschrig}},
  \bibinfo{journal}{Phys. Rev. B } \textbf{\bibinfo{volume}{59}},
  \bibinfo{pages}{1743} (\bibinfo{year}{1999}). For details of the calculations
see Ref.~\onlinecite{rosner02}.

\bibitem[{\citenamefont{Zhernov and Drechsler}(1985)}]{zhernov85}
\bibinfo{author}{\bibfnamefont{A.~P.} \bibnamefont{Zhernov}} \bibnamefont{and}
\bibinfo{author}{\bibfnamefont{S.~L.} \bibnamefont{Drechsler}},
  \bibinfo{journal}{Fizika Nizkikh Temperatur (Sov. Low Temp. Phys.)} \textbf{\bibinfo{volume}{11}},
  \bibinfo{pages}{899 (495)} (\bibinfo{year}{1985}).


\end{thebibliography}

\end{document}